\documentstyle[prl,aps,preprint]{revtex}
\begin{document}

\title{Magnetoresistance of metallic perovskite oxide LaNiO$_{3-\delta}$} 
\author{N.~Gayathri\thanks{gayathri@physics.iisc.ernet.in} and
A.~K.~Raychaudhuri\thanks{On lien to National Physical Laboratory, New Delhi 
110012, India.}} 
\address{Department of Physics, Indian Institute of Science,
Bangalore 560012, India}
\author{X.~Q.~Xu, J.~L.~Peng and R.~L.~Greene}
\address{Centre for Superconductivity Research, 
University of Maryland, \\ College Park, Maryland 20742, USA}
\maketitle
\begin{abstract}
We report a study of the magnetoresistance (MR) of the metallic
perovskite oxide LaNiO$_{3-\delta}$ as a function of the oxygen
stoichiometry $\delta$ ($\delta \leq$ 0.14), magnetic field ( H $\leq
6T$ ) and temperature (1.5K $\leq $ T $\leq $ 25K).  We find a strong
dependence of the nature of MR on the oxygen stoichiometry. The MR at
low temperatures change from positive to negative as the sample
becomes more oxygen deficient ( i.e, $\delta$ increases).  Some of
the samples which are more resistive, show a resistivity minima at
$T_{min}$ $\approx$ 20K. We find that in these samples the MR is
positive at T > $T_{min}$ and negative for T < $T_{min}$. We conclude
that in the absence of strong magnetic interaction, the negative MR in
these oxides can arise from weak localisation effects. 
\end{abstract}

\newpage
\section{Introduction}
In recent years the electrical conduction in transition metal
perovskite oxides at low temperature has attracted considerable
research interest\cite{AKR}. Most of these oxides studied belong to
the ABO$_3$ class or its closely linked derivatives. In recent years
the observation of colossal magnetoresistance (CMR) in rare-earth
maganates belonging to the ABO$_3$ class has added yet another new
phenomena \cite{CHAA}. These oxides have chemical formula of the type
La$_{1-x}$(Sr/Ca)$_{x}$MnO$_{3}$ with $x\approx 0.2-0.4$. The
colossal {\bf negative} MR seen in these oxides have become a topic
of intense research activity. In these mixed valent manganates (doped
to create Mn$^{4+}$) the CMR is closely linked to the occurence of
strong ferromagnetism which arises from the Double Exchange (DE)
interactions. The additional important contribution to the CMR
phenomenon comes from the strong Jahn-Teller effect of the
Mn$^{3+}$ ions. In a closely related compound
La$_{1-x}$Sr$_{x}$CoO$_{3}$ (with Co$^{3+}$ and Co$^{4+}$) which is
also a ferromagnetic metallic oxide for $x\geq 0.2$ the MR was found
to be much smaller except in the composition range $x$ $\approx$ 0.2
where the oxide undergoes a composition driven metal-insulator
transition \cite{MAHI2}. The negative MR in this material for $x$
$\approx$ 0.2 is believed to be related to the formation of
ferromagnetic clusters. In both these oxides the electron is on the
verge of getting localized. On the other hand there are metallic
oxides also belonging to ABO$_3$ class like Sr$_{x}$NbO$_{3}$ (x$\geq$
0.75) \cite{SRNB} and Na$_{0.9}$WO$_{3}$ \cite{NG} with relatively
high conductivity which have positive MR at all temperatures. In
these oxides the MR has a quadratic dependence on the applied field
$H$ much like a band metal. The principal differences between the
two classes of perovskite oxides seems to be the extent of electron
localization and the presence of ferromagnetic interactions. This
leads us to the following interesting question: {\it If we start with
a metallic perovskite oxide showing positive MR and gradually make
the electrons localized (by some means), will the positive MR change
to a negative MR as the electrons get localized?} We investigate this
question in the metallic perovskite LaNiO$_{3-\delta}$ in which we
find that the MR changes from positive to negative as the material is
made more resistive by increasing $\delta$. LaNiO$_3$ belonging to
the ABO$_3$ class is structurally similar to LaCoO$_3$ and LaMnO$_3$.
However unlike the cobaltate and the manganate, it is a fairly good
metal with low resistivity \cite{NYV,DESHI,RLG,REGI,SANCH}. However,
as the oxygen deficiency is increased the resistivity is enhanced.
Eventually the material becomes an insulator at $\delta$ > 0.25
\cite{SANCH}. The investigation of the MR in this oxide is thus of
importance because it allows us to investigate the MR in a class of 
disordered oxides where a change in oxygen stoichiometry leads to a
qualitative change in the nature of electronic transport without
bringing in strong magnetic effects atleast for $\delta \leq$ 0.2.

LaNiO$_3$ is an interesting material because of its potential in
application as metallic interconnects (or electrodes) in thin film
oxide electronics particularly those needing epitaxial multilayer
perovskite oxide films \cite{HEGDE}. The MR data of this oxide will
thus be of importance technologically.

\section{Experimental details}
The samples of LaNiO$_{3-\delta}$ were prepared from co-precipitation
and subsequent decomposition of nitrates of La and Ni as described in
refs.~[8] and [12]. The samples were heat treated
under different oxygen atmosphere to get varying oxygen content and
the oxygen stoichiometry $\delta$ was determined by iodometric
titration. The grain sizes as measured by scanning electron
microscope were typically of the order of 2 $\mu$m and there was no
variation in the grain sizes of the samples with different $\delta$.
[See ref [13] for details of sample preparation and charactrisation].

The electrical resistivity $\rho$(T) and the MR were measured by a
high precision low frequency (20 Hz) ac bridge technique with a
precision of $\pm$10ppm.  The resistivity was measured in the range
0.4K -300K and the MR was measured in the range 1.5K- 30K in a field of
up to 6 Tesla. 

\section{Results and Discussion}
In figure~1 we present the resistivity measurements for the three
samples with the general formula LaNiO$_{3-\delta}$.  We identify the
samples in decreasing order of $\delta$ and have called them A,B and
C for $\delta$ = 0.14, 0.08 and 0.02 respectively. For all the
samples the $\rho$(T) curve show a metallic behaviour although the
residual resistivity $\rho_{0}$ increases by a factor of 30 as
$\delta$ changes from 0.02 to 0.14. We restrict our investigations to
$\delta\leq$ 0.14 to avoid any complications arising from magnetic
clusters which appear for $\delta \geq$ 0.2 \cite{SANCH}.

For the sample with $\delta$=0.14 which has the highest $\rho_0$ the
mean free path ($\ell_{el}$), estimated from the simple free electron
approximation, is somewhat smaller than the Ni-O bond length
($\approx$ 2\AA). Interestingly, it can be seen that for this sample
there is a resistivity minima around 20K. This is very similar to the
behavior of highly disordered metallic alloys.

The exact temperature dependence of $\rho$(T) in LaNiO$_{3-\delta}$
with varying $\delta$ has been critically looked into \cite{GAYA}. We
do not discuss these in detail here. We just mention that in the samples
B and C which show no resistivity minima we have a power law dependence
of the resistivity with T given as \cite{GAYA}:

\begin{equation}
\rho(T) = \rho_0 + \alpha T^{1.5} + \beta f(T)  \label{eq:fullt}
\end{equation}

\noindent where $f$(T) is the function giving a linear temperature
dependence at high temperature (T~>~200K) and a T$^n$ (n$\geq$ 2)
dependence for low temperatures. $\rho_0$ is the residual resistivity
and $\alpha$ and $\beta$ are constants. For the sample A, which shows
the highest resistivity, we have a resistivity minima
at T$_{min}$~=~20K and $\rho$(T) for T~<~T$_{min}$ shows a shallow
rise (see figure~1). We find that this shallow rise of $\rho$ below
T$_{min}$ can be best accounted for by using expressions for weak
localization (WL). We also find below that the MR measurements
supports this view. We quantify the weak localization
\cite{TVR} contribution by an additional term -$\gamma$T$^{0.75}$
which is incorporated into eqn.\ref{eq:fullt}. Thus, we fit the
resistivity of sample A using the expression

\begin{equation}
\rho = \rho_0 - \gamma T^{0.75} + \alpha T^{1.5} + \beta f(T) 
\label{eq:afullt}
\end{equation}

\noindent We find that this relation fits the resitivity curve with an error
better than $\pm$ 0.5\% over the entire temperature range. The fits
for the samples are shown as a solid line in figure~1.

In figure~2 we show the MR of the three samples as a function of the
applied field H at T = 4.2K. We define the MR as
$\Delta\rho$/$\rho_{0H}$ = ($\rho$(H) - $\rho$(0))/$\rho_{0H}$, where
$\rho_{0H}$ for the zero field resistivity at a given temperature T.
The magnetoresistance of the LaNiO$_{3-\delta}$ system is small but
has a distinct dependence on the oxygen stoichiometry. It is positive
for the least resistive sample C, it is negative for sample A which
has the highest resistivity and it is almost zero (< 0.05\%) for
sample B with intermediate resistivity. The MR thus becomes
progressively negative as the resistivity increases.

The field dependence of MR also changes as $\delta$ increases. For
sample C, MR has a quadratic dependence on H which is shown in figure~3.
The MR seems to follow K\"ohler rule \cite{KHOLER}. The MR for this
sample is positive at all T. According to Koehler rule the dependence
of MR on temperature, magnetic field and purity of the sample should
be a function of H$\tau$ only, where $\tau$ is the relaxation time.
Since $\rho_{0H} \propto
\tau^{-1}$, $\Delta\rho/\rho_{0H} \propto$ $\it{k}$(H/$\rho_{0H}$),
where $\it{k}$(H/$\rho_{0H}$) is a function of H/$\rho_{0H}$.  For
normal metals the MR shows a quadratic dependence with field due to
the band MR. Hence we can write:

\begin{eqnarray}
\frac{\Delta\rho}{\rho_{0H}} = \beta_\rho\left(\frac{H}{\rho_{0H}}\right)^2 
\label{eq:kho}
\end{eqnarray}

\noindent where, $\beta_\rho$ is a constant for the material. Fitting
the MR of sample C with the above expression we get $\beta_\rho$ =
3.98$\times$10$^{-12}$ ($\Omega$cm/T)$^2$. For normal metals the value of
$\beta_\rho \approx$ 10$^{-15}$ ($\Omega$cm/T)$^2$. This difference
is very interesting because if the positive MR indeed arises from a
band mechanism then it is one of the largest MR arising from the band
mechanism.  It will be of interest to compare the MR of this oxide
with other oxides showing positive MR \cite{SRNB,NG}. The other known
metallic oxides showing positive MR, Sr$_{x}$NbO$_3$ (x $\leq$ 0.8)
\cite{SRNB} and Na$_{0.9}$WO$_3$ \cite{NG} shows a H$^2$ dependence
over a small field range (H $\leq$ 3 Tesla). The values of
$\beta_\rho$ for Sr$_{0.85}$NbO$_3$ ($\rho_{4.2K}$ = 1.75m$\Omega$cm)
and Na$_{0.9}$WO$_3$ ($\rho_{4.2K}$ = 3$\mu\Omega$cm) are
1.127$\times$10$^{-9}$ ($\Omega$cm/T)$^2$ and 4x10$^{-15}$
($\Omega$cm/T)$^2$ respectively.  It seems therefore in these oxides
the constant $\beta_\rho$ increases as $\rho$ increases. This
observation needs more investigation to establish whether this indeed
is a general feature of metallic oxides.

For sample A the MR is not only negative, the field dependence is
also qualitatively different. We will show below that it is possible
to explain the field dependence of the negative MR as arising from
destruction of weak localisation on application of magnetic field.
For sample B, the MR being almost zero we didn't attempt any fit.
Most likely for this sample we have almost equal contribution from
both the positive and negative terms.

The temperature dependence of the MR also carries a distinct mark of
the oxygen stoichiometry. As shown in figure~4, the MR of sample A is
negative for T < T$_{min}$ and positive for T > T$_{min}$, where
T$_{min}$ is the temperature of resistivity minima (see figure~1). It
has been discussed before that the resistivity minima can arise due
to the contribution by WL for T < T$_{min}$. For T > T$_{min}$ the MR
is positive as in sample C. For T < T$_{min}$ the WL is supressed by
the application of the magnetic field and hence there will be a
negative contribution to the MR. To look at this quantitatively we
have fitted the magnetoconductance curves to the WL expression
\cite{KAWA,BAX} with an additional H$^2$ term to account for the
K\"ohler type negative magnetoconductance (or positive MR). In the
following we used conductivity change ($\delta\sigma$) instead of
resistivity change ($\delta\rho$) to follow the customary procedure
for WL expressions.  The expression thus used was \cite{BAX}:

\begin{eqnarray}
\delta\sigma =
\frac{e^2}{2\pi^2\hbar}{\left(\frac{eH}{\hbar}\right)}^{1/2} 
{\cal H}\left(\frac{4{\cal D}\tau_\phi eH}{\hbar}\right) -
\beta_\sigma H^2
\label{eq:mr} 
\end{eqnarray}

\noindent where
\begin{eqnarray}
{\cal H}(x) = 2\left[\sqrt{2+\frac{1}{x}}-\sqrt{\frac{1}{x}}\right]
-\left(\frac{1}{2}+\frac{1}{x}\right)^{-1/2} -
\left(\frac{3}{2}+\frac{1}{x}\right)^{-1/2} + 
\frac{1}{48}\left(2.03+\frac{1}{x}\right)^{-3/2} \nonumber
\end{eqnarray}

\noindent Here, ${\cal D}$ is the electron diffusivity, $\tau_\phi$
is the electron phase relaxation time and $\beta_\sigma$ is a
constant. In this expression $\beta_\sigma$ and $\cal D\tau_\phi$ can
be free parameters. However, to restrict the number of free
parameters we fixed the parameter $\beta_\sigma$ following K\"ohler's
rule using the following strategy.

The K\"ohler's rule (see eqn \ref{eq:kho}) expressed in terms of the
conductivity is given by $\delta\sigma_{Kohler}$ = -$\beta_\sigma$
H$^2$ where $\beta_\sigma$ = $\beta_\rho/\rho_{0H}^3$. Assuming the
validity of the K\"ohler's rule for the samples A and C would imply
that both samples will have the same $\beta_\rho$. Thus, using
$\beta_\rho$ as found from sample C in figure~3 we have estimated the
value of $\beta_\sigma$ for sample A. The values of $\beta_\sigma$
for T $\leq$ 4.2K are $\approx$ 0.01 (S/cm)/T$^2$. Using these
$\beta_\sigma$ we fit the magnetoconductance data to eqn
\ref{eq:mr} using $\cal D\tau_\phi$ as the only fit parameter. A good
fit can be obtained for the magnetoconductance data as shown in the
inset of figure~4. The $\cal D\tau_\phi$ values (fit parameters) for the
two temperatures were obtained as 1.26x10$^{-12}$ cm$^2$ for T=2.3K
and 1.12x10$^{-12}$ cm$^2$ for T=4.2K.  Using the above values of
$\cal D\tau_\phi$, we have estimated the value of the WL correction
to the zero field conductivity of the sample. We get
$\delta\sigma_{WL}$ ($\approx (e^2/2\pi^2\hbar) (1/\sqrt{\cal
D\tau_\phi})$ to be about 12.3 S/cm which is about a factor of 2.5
more than that estimated from the resistivity data of this sample at
low temperature ($\approx$ 4.8 S/cm). Within the limits of
experimental error we can say that this is a good agreement since,
the estimate of $\delta\sigma_{WL}$ from the resistivity involves
some uncertainities because of the presence of other terms with much
larger contributions (see eqn.~2). Also we have only an approximate
value for $\beta_{\sigma}$. We can thus conclude that the negative MR
seen in this oxide can arise from the weak localization contribution.
At T~=~25~K the positive contribution wins over. But given the
smallness of MR it is difficult to obtain any meaningful fit.

One important effect that we should address is the formation of
local magnetic moments and any ferromagnetic interaction arising from
it which can give rise to negative MR. 
While stoichiometric LaNiO$_{3}$ is a Pauli paramagnet,
creation of Ni$^{2+}$ ions in oxygen deficient samples can lead to a
Curie-Weiss type contribution. We have measured the susceptibility for
sample A. The susceptibility and the analysis are reported in ref.[13]. 
We mention the important results, since they are relevant to this paper. 
For T<100K we could fit the susceptibility to $\chi$ = $\chi_0$+
C/(T-$\theta$). A fit to the 
expression gives $\chi_0$ = 5.95x10$^{-6}$ emu/gm, $\theta$ = -51~K
and C = 3.83x10$^{-4}$ emu~ K/gm. The large value of C can arise from
the Ni$^{2+}$ formation as $\delta$ increases \cite{GAYA}.  For
sample A, for which $\delta$ = 0.14, about 30\% of the Ni in the
system is in the Ni$^{2+}$ state. However, not all of the Ni$^{2+}$
carry moments because part of the Ni$^{2+}$ which are in square
planar cordination are diamagnetic. Though there is a net small
antiferromagnetic interaction ($\theta \approx$ -51~K), the
susceptibility shows no sign of any magnetic ordering or cluster
formation. It can be contrasted to the composition LaNiO$_{2.75}$
which shows the transition to an insulating state and a clear
signature of a spin-glass or cluster glass like transition near 150K
\cite{SANCH}. From these observations we conclude that the magnetic
interactions will not play an important role for the MR of the
samples ($\delta \leq$ 0.14) which we have investigated. It may be
that for $\delta \geq$ 0.2 the cluster formation can lead to MR
dominated by magnetic interaction effects.

In summary, we have measured the magnetoresistance of the metallic
perovskite oxide LaNiO$_{3-\delta}$ as a function of the oxygen
defect $\delta$. The most metallic sample has a clear positive MR
which progressively goes to the negative magnitude as the oxygen
defect increases. The positive MR seems to arise from a band-like
contribution as seen in normal metals but the magnitude is much
larger than that seen in conventional metals. The negative MR seen in
the oxygen deficient sample arises from weak-localisation
contribution and is not due to any magnetic interactions. Our
experiment shows that the effects like weak-localisation of electrons
can lead to small but finite negative MR in these oxides which may be
masked by stronger negative MR arising from magnetic origin in
other oxides.

\vspace{2 truecm}
{\bf Acknowledgements:} One of us (AKR) wants to thank the Department
of Science and Technology, Government of India for a sponsored
project.  NG wants to thank the Council for Scientific and Industrial
Research for a fellowship. RLG acknowledges the supported by NSF
under Grant No: DMR9510475.

\newpage
\centerline{FIGURE CAPTIONS}
\noindent{\bf Figure 1} Resistivities ($\rho$) of the three samples
of LaNiO$_{3-\delta}$. The samples A, B \& C have $\delta$ = 0.14,
0.08 and 0.02 respectively. The line shows the fit discussed in the
text (see eqns:\ref{eq:fullt} and \ref{eq:afullt}).

\noindent{\bf Figure 2} Magnetoresistance(MR) as a function of field
(H) for the three samples A,B and C at T=4.2K. $\rho_{0H}$ is the zero
field resistivity.

\noindent{\bf Figure 3}MR as a function of H$^2$ for sample C at T=4.2K.

\noindent{\bf Figure 4} Magnetoresistance as a function of field for Sample A
for three different temperatures 2.3K and 4.2K (T<T$_{min}$) and
25K (T>T$_{min}$). {\bf Inset:} Fit of the data to the MR relation
for weak localisation for T<T$_{min}$ (plotted as conductivity
$\delta\sigma$)

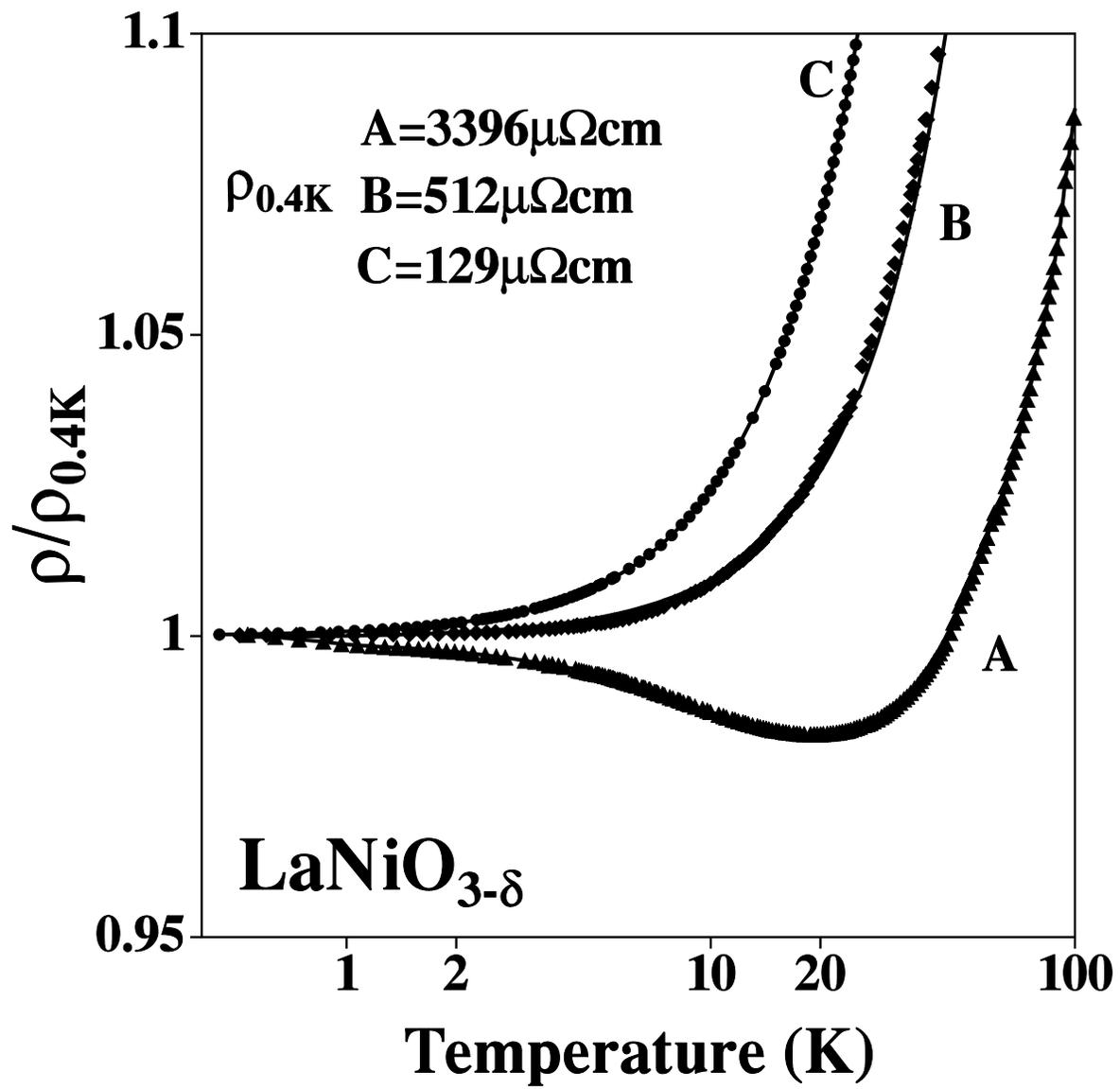
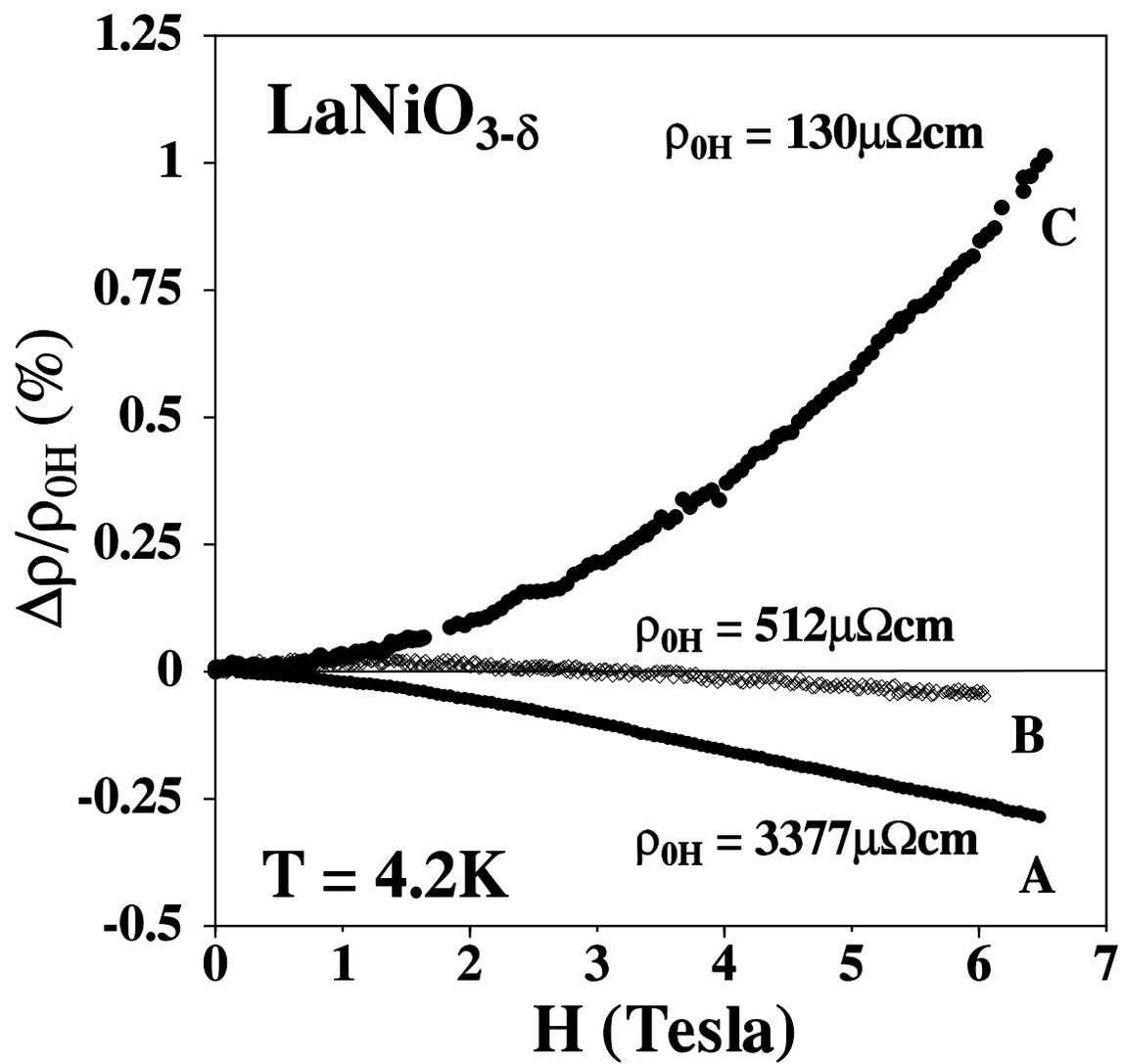
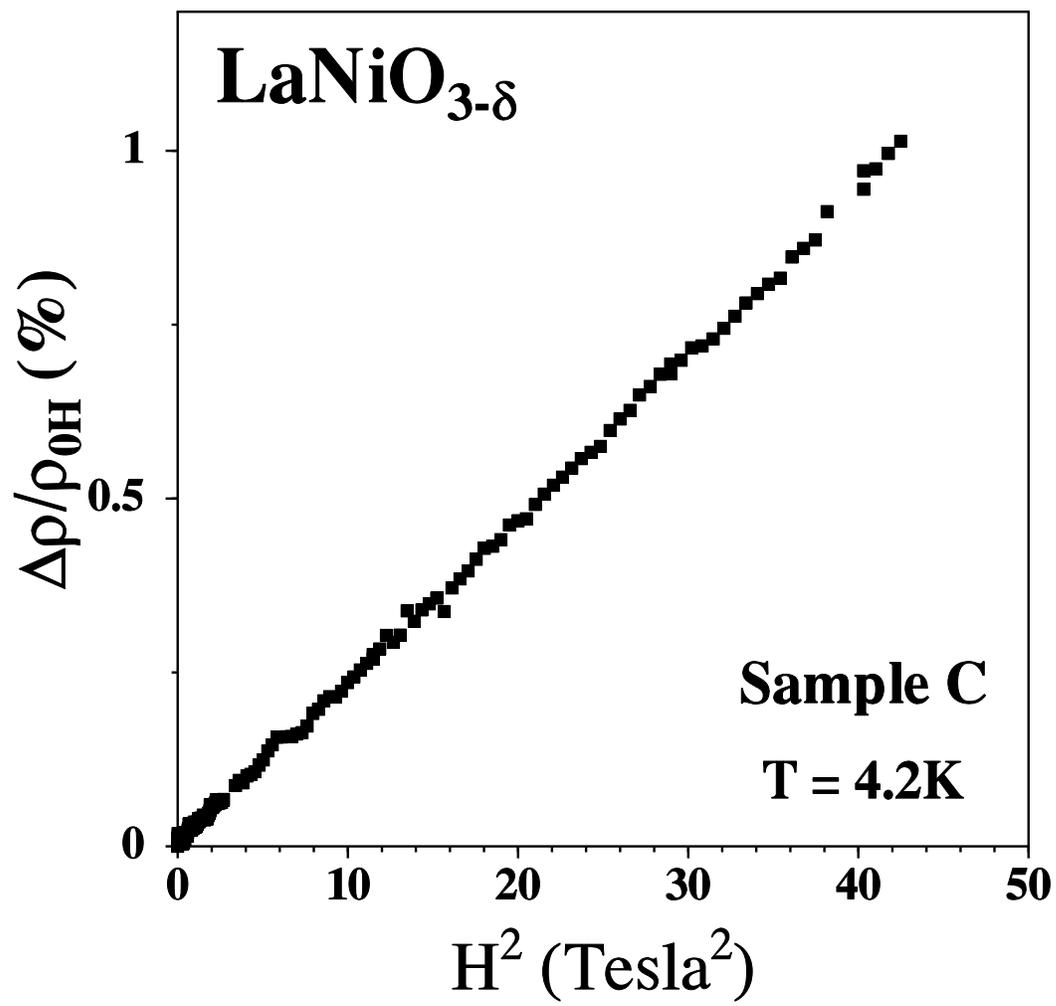
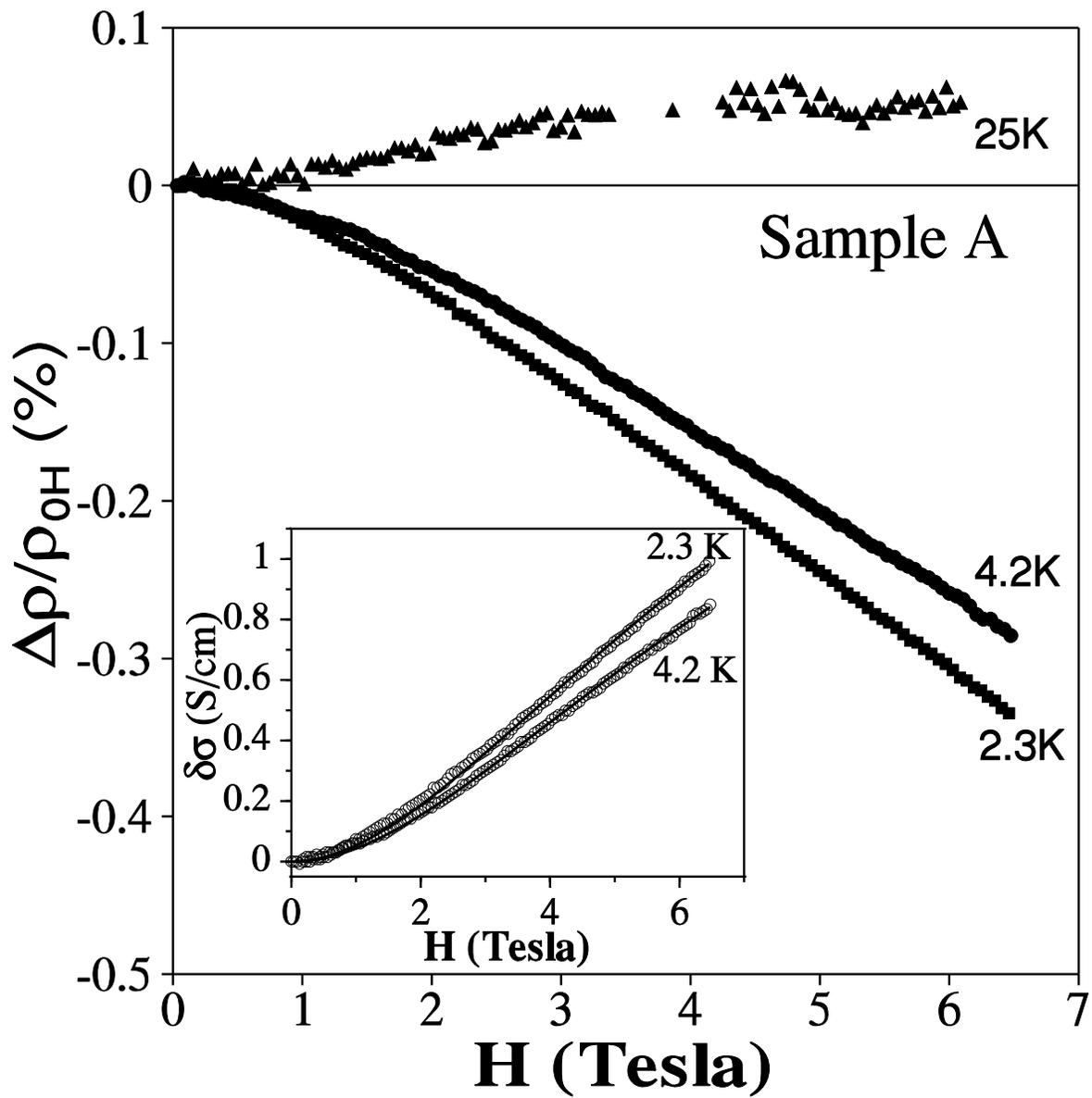
\end{document}